%% file: main.tex
\documentclass[a4paper,UKenglish]{lipics}

\usepackage{microtype}

\usepackage{xspace}
\usepackage{graphicx}

\usepackage{bold-extra}
\usepackage{listings}

\usepackage{fancybox}

\lstdefinelanguage{JavaScript}{
  keywords={break, case, catch, continue, debugger, default, delete, do, else, finally, for, function, if, in, instanceof, new, return, switch, this, throw, try, typeof, var, void, while, with},
  morecomment=[l]{//},
  morecomment=[s]{/*}{*/},
  morestring=[b]',
  morestring=[b]",
  sensitive=true
}

\makeatletter
\newenvironment{centeredbox}{
\begin{Sbox}}{
\end{Sbox}\centerline{\parbox{\wd\@Sbox}{\TheSbox}}}
\makeatother

\title{Simple and Effective Type Check Removal through Lazy Basic Block Versioning}
\author[1]{Maxime Chevalier-Boisvert}
\author[2]{Marc Feeley}
\affil[1]{DIRO, Universit\'e de Montr\'eal\\
  Montr\'eal, QC, Canada\\
  \texttt{chevalma@iro.umontreal.ca}}
\affil[2]{DIRO, Universit\'e de Montr\'eal\\
  Montr\'eal, QC, Canada\\
  \texttt{feeley@iro.umontreal.ca}}
\authorrunning{M. Chevalier-Boisvert and M. Feeley}
\Copyright{Maxime Chevalier-Boisvert and Marc Feeley}


\volumeinfo{John Tang Boyland}
{1}
{29th European Conference on Object-Oriented Programming (ECOOP'15)}
{37}
{1}
{999}
\EventShortName{ECOOP'15}%
\DOI{10.4230/LIPIcs.ECOOP.2015.999}
\serieslogo{}

\begin{document}

\maketitle

\input{abstract}

\section{Introduction}\label{sec:intro}
\input{introduction}

\section{Basic Block Versioning}\label{sec:versioning}
\input{versioning}

\section{Implementation in Higgs}\label{sec:implementation}
\input{implementation}

\section{Evaluation}\label{sec:evaluation}
\input{evaluation}

\section{Related Work}\label{sec:related}
\input{related}

\section{Limitations and Future Work}\label{sec:future}
\input{future}

\section{Conclusion}\label{sec:conclusion}
\input{conclusion}

\section*{Acknowledgements}
Special thanks go to Paul Khuong, Laurie Hendren, Erick Lavoie, Tommy Everett,
Brett Fraley and all those who have contributed to the development of Higgs.

This work was supported, in part, by the Natural Sciences and Engineering
Research Council of Canada (NSERC) and Mozilla Corporation.

\bibliographystyle{plain}
\bibliography{main}

\end{document}

%% file: abstract.tex
\begin{abstract}

%

Dynamically typed programming languages such as JavaScript and Python defer
type checking to run time. In order to maximize performance,
dynamic language VM implementations must attempt to eliminate redundant dynamic
type checks. However, type inference analyses are often costly and involve
tradeoffs between compilation time and resulting precision. This has lead to
the creation of increasingly complex multi-tiered VM architectures.

This paper introduces {\em lazy basic block versioning}, a simple JIT compilation
technique which effectively removes redundant type checks from critical code
paths. This novel approach lazily generates type-specialized versions of basic blocks
on-the-fly while propagating context-dependent type information. This
does not require the use of costly program analyses, is not restricted by the
precision limitations of traditional type analyses and avoids the
implementation complexity of speculative optimization techniques.

We have implemented intraprocedural lazy basic block
versioning in a JavaScript JIT compiler. This approach is compared with a
classical flow-based type analysis. Lazy basic block versioning performs as
well or better on all benchmarks. On average, \input{testredumv5}\unskip\% of type tests are
eliminated, yielding speedups of up to \input{maxspeedupmv5}\unskip\%.
We also show that our implementation generates more efficient machine code than
TraceMonkey, a tracing JIT compiler for JavaScript, on several benchmarks.
The combination of implementation simplicity, low algorithmic complexity
and good run time performance makes basic block versioning attractive for
baseline JIT compilers.

\end{abstract}

%% file: testredumv5.tex
71

%% file: maxspeedupmv5.tex
50

%% file: introduction.tex
%
%


A central feature of dynamic programming languages
is that they defer type checking to run time.
In order to maximize performance, efficient implementations of dynamic
languages seek to type-specialize code so as to eliminate dynamic type checks
when possible. Doing so requires proving that these type checks are unnecessary
and generating type-specialized code.

Traditionally, the main approach for eliminating type checks has been to use
type inference analyses. This is problematic for modern dynamic languages such
as JavaScript and Python for three main reasons. The first is that these languages are
generally poorly amenable to whole-program type analyses. Constructs such as
{\tt eval} and dynamic loading of modules can destroy previously valid type
information. The second is that these analyses can be expensive in
terms of computation time and memory usage, making them unsuitable for
JIT compilers, particularly baseline compilers. To reduce analysis cost, it is
often necessary to sacrifice precision. A last issue is that some type checks
simply cannot be eliminated through analysis alone, without code transformations.


Because dynamic programming languages are generally poorly amenable to
type inference, and whole-program analyses are often too
expensive for JIT compilation purposes, state of the art JavaScript VMs such
as SpiderMonkey, V8 and JavaScriptCore rely on increasingly complex
multi-tiered architectures integrating interpreters and
multiple JIT compilers with different optimization capabilities (baseline
compilers to aggressively optimizing compilers). At the
highest optimization levels, modern JIT compilers typically make use of type feedback, type inference
analysis and also speculative optimization and deoptimization~\cite{mozti} with On-Stack
Replacement (OSR).

We introduce a simple approach to JIT compilation that
generates efficient type-specialized code without the use of costly type
inference analyses or type profiling. The approach, which we call
lazy basic block versioning, lazily clones and specializes basic blocks
on-the-fly in a way that allows the compiler to accumulate type information
while machine code is generated, without a separate type analysis pass.
The accumulated information allows the removal of redundant type tests,
particularly in performance-critical paths.

Lazy basic block versioning lets the execution of the program itself drive the
generation of type-specialized code, and is able to avoid some of the
precision limitations of traditional, conservative type analyses as well as
avoiding the implementation complexity of speculative optimization techniques.

This paper relates our experience implementing lazy basic block versioning and
reports on its effectiveness as a code generation technique. The rest of the
paper is organized as follows. Section~\ref{sec:versioning} explains the basic
block versioning approach, comparing it with the related approaches of static
type analysis and trace compilation. Section~\ref{sec:implementation}
describes an implementation within Higgs, an experimental JIT compiler for
JavaScript. Section~\ref{sec:evaluation} presents an evaluation of the
performance of this implementation. Related work is presented in
Section~\ref{sec:related}.

%% file: versioning.tex
In the basic block versioning approach, the code generator maintains a
typing context (or type map) which indicates what is known of the type
of each live local variable at the current program point. All local variables start out
with the {\em unknown} type at function entry points. While generating code, the code
generator updates the typing context by inferring the result type of data
operations it encounters. Conditional branch instructions corresponding to type tests
create two new typing contexts for outgoing branch edges. In each context, a type
is assigned to the variable being tested (either the type tested or {\em unknown}).
When a type test branch instruction
is encountered and the type of the argument is known, the branch direction can
be determined at code generation time and the type test eliminated.

The compiler may generate code for multiple instances of a given basic block;
one version for each typing context encountered on a branch to
that basic block.  This allows specializing the basic block and its successors
by taking the types of live variables into account. While basic block
versioning works at the level of individual basic blocks, the propagation of
typing contexts to successor blocks allows type-specializing entire control
flow graphs.

An important difference between this approach and traditional type
analyses is that basic block versioning does not compute a fixed point
on types to be inferred. Variables which may have multiple different
types at the same program point are handled more precisely with basic
block versioning due to the duplication of code.  In a traditional
type analysis, the union of several possible types would be assigned
to such variables, causing the analysis to be conservative.  With
basic block versioning, distinct basic block versions, and thus distinct code
paths, will be created for each type previously encountered, allowing a
precise context-dependent tracking of types.

With basic block versioning, loops in the control flow graph need not be
handled specially.  A first version of the loop header is generated for a
typing context $C_1$.  At the point(s) where control flow branches back to the
loop header, a new version of the loop may be generated if the typing context
$C_2$ is different from $C_1$.  Given that the number of possible contexts is finite,
a fixed point is eventually reached, that is, the typing context at branches
to the loop header will eventually match one of $C_1$, $C_2$, \ldots, $C_N$.
The number of versions actually generated is expected to be low because the
type of most variables remains stable for the duration of a function.

There is a risk of a combinatorial explosion when multiple versions of
basic blocks are created eagerly.  Consider the simple statement {\tt
  x=a+b+c+d}.  If the types of {\tt a}, {\tt b}, {\tt c} and {\tt d}
are {\em unknown}, and those variables are live after the assignment, and there are two possible numerical types
({\tt int} and {\tt float}), there could be up to 16 versions of the basic block containing
the assignment to {\tt x}, one version for each set of type assignments
to the variables being summed.  In general, if basic block versioning is
performed in an {\em eager} fashion and there are $t$ possible types of values
and a function has $v$ variables, then there can be up to $t^v$ versions of
some basic blocks in that function.  However, the logic of a program puts
constraints on possible type combinations. In practice, not all the
combinations of types are observed during an execution of a program.

It is often the case that variables are monomorphic in type (i.e.~they always contain
the same type of value). We can take advantage of this by {\em lazily} creating
new block versions on demand. Versions for a particular context are only
generated when that context is encountered
during execution.  {\em Lazy basic block versioning} doesn't completely eliminate the
possibility of a combinatorial explosion in pathological cases, but
this can be prevented by placing a hard limit on the number of
versions generated for any given block.  Some increase in code size is to be expected, but no
more than a constant factor.  Mueller and Whalley have
shown~\cite{code_repl_uncond} that specializing code through
replication, while increasing the static size of machine code, can
reduce the dynamic count of executed instructions and result in better
cache usage.

Traditional type analyses often cannot infer a type for a variable,
either because there is insufficient semantic information in the source
program, or because the analysis is limited in its capabilities. For example,
with an intraprocedural type analysis of JavaScript, no type information is
known about function parameters. Without transforming the program, many
variable types cannot be recovered by analysis alone. Moreover, the {\em unknown}
type may propagate through primitive operations and effectively poison the
results of such type analyses.

As will be demonstrated in Section \ref{sec:evaluation}, a key advantage of
basic block versioning over program analyses lies in its ability to {\em recover
unknown types}. The versioning approach is able to exploit type tests that are
implicitly part of the language semantics to gain type information, and then
generate new block versions where the additional type information remains
known. Basic block versioning automatically unrolls some of the
first iterations of loops in such a way that type tests are hoisted out of
loop bodies. For example, if variables of {\em unknown} type are used
unconditionally in a loop, their type will be tested only in the first
iteration of the loop. The type information gained will allow further
iterations to avoid redundant type tests.

Lazy basic block versioning bears some similarity to trace
compilation~\cite{dynamo} in the use of code duplication and type
specialization to eliminate type tests~\cite{trace_spec}.  Trace compilation typically relies on
an interpreter to detect hot loops and record traces. It is also most effective
on loop-heavy code.  In contrast, lazy basic block versioning can handle any
code structure just as effectively. It avoids the dual language
implementation (interpreter and trace compiler) and requires no special
infrastructure for profiling or recording traces.

The relative simplicity of tracking typing contexts and previously generated
basic block versions means that the compiler avoids algorithms of high
computational complexity. With a hard limit on the number of block versions,
code generation time and code size scale linearly with
the size of the input program. Lazy basic block versioning requires no external
optimization or analysis passes to generate type-specialized code. This makes
the approach interesting for use in baseline JIT compilers.

%% file: implementation.tex

We have implemented lazy basic block versioning inside a JavaScript virtual machine
called Higgs. This virtual machine comprises a JIT compiler targeted at x86-64
POSIX platforms. The current implementation of Higgs supports most of the
ECMAScript 5 specification~\cite{js_spec}, with the exception of the
{\tt with} statement and the limitation that {\tt eval} can only access global
variables, not locals. Its runtime and standard libraries are self-hosted,
written in an extended dialect of ECMAScript with low-level primitives. These
low-level primitives are special instructions which allow expressing type
tests as well as integer and floating point machine instructions in the source
language.

In Higgs, functions are parsed into an abstract syntax tree and lazily
compiled to a Static Single Assignment (SSA) Intermediate Representation (IR)
when they are first called. Inlining is performed at this time according to
simple fixed heuristic rules. Specific JavaScript runtime functions
including arithmetic, comparison and object property access 
primitives are always inlined. This inlining allows exposing type tests and
typed low-level operations contained inside primitives to the backend, which
implements basic block versioning.

A basic block version corresponds to
a basic block and an associated context containing type information about live
values at the start of the block. Machine code generation always begins with
the function's entry block and a default entry context being queued for
compilation. Typing contexts in Higgs are implemented as sets of pairs
associating live SSA values to unique type tags (see Section
\ref{sec:type_tags}). Values for which no type information is known do not
appear in the set. As each instruction in a block is compiled, information
is both retrieved from and inserted into the current context. Information
retrieved may be used to optimize the compilation of the current instruction
(e.g.\ eliminate type tests). Instructions will also write their own output
type in the context if known.

\begin{figure}[!tb]
\begin{lstlisting}[language=javascript]
/**
Context compatibility test function:
- Perfectly matching contexts produce score 0
- Imperfect matches produce a score > 0
- Incompatible matches produce Infinity
*/
Number contextComp(Context predCtx, Context succCtx)
{
    Number score = 0;

    // For each value live in the successor
    foreach (value in succCtx)
    {
        auto predType = predCtx.getType(value);
        auto succType = succCtx.getType(value);

        // If the successor has no known type,
        // we would lose a known type
        if (predType != UNKNOWN &&
            succType == UNKNOWN)
            score += 1

        // If the types do not match,
        // contexts are incompatible
        else if (predType != succType)
            return Infinity;
    }

    return score;
}
\end{lstlisting}
\caption{Context compatibility test function\label{fig:ctxComp}}
\end{figure}

To guard against pathological cases where an unreasonably large number of
versions would be generated, we have added one tunable parameter,
{\tt maxvers}, which specifies the maximum number of specialized
versions that can be generated for any given basic block. Before the limit for a
given block is reached, requests for new versions matching an incoming
context will either find an existing exact match for the context, or
compile a new version matching the incoming context exactly. Once the
limit is reached for a particular block, requests for new versions of this
block will first try to find an inexact but compatible match for
the incoming context. An existing version is compatible
with the incoming context if the value types assumed by the existing
version are a subset of those specified in the incoming context.

The context compatibility test is shown in Figure
\ref{fig:ctxComp}. A context containing less constraining types than
the incoming context is compatible, but one that has more constraining
types than the incoming context is not. Essentially, this allows for the
loss of type information when transitioning along control flow edges. If the
version limit was reached and no compatible match is found for a given
block, a fully generic version of the block that
assigns the {\em unknown} type to all live variables will be generated.
This generic version
is compatible with all possible incoming contexts. When the {\tt maxvers} parameter is set to
zero, basic block versioning is disabled, and only one generic version
of each basic block may be generated.

\subsection{Lazy Code Generation}

Limiting the number of versions generated by {\em eager} basic
block versioning to avoid combinatorial code growth is a difficult problem.
Simply imposing a hard version limit is not a satisfactory solution because it is
nontrivial to determine ahead of time which typing contexts are more probable
than others, and which may not occur at all. This is particularly problematic in
a JIT compiler, since compiling versions for type combinations that will not
occur at run time translates into wasted compilation time, code bloat
and poor usage of the instruction cache. There is also the issue of ordering machine code in
memory so as to minimize the number of branches taken.

Clearly, basic block versioning ought to be guided by run time types,
but gathering profiling data using traditional means could be expensive.
Furthermore, the resulting data may be large and complex to analyze. Instead,
Higgs delays the generation of block versions
and lets the run time behavior of programs drive this process. The
{\em execution of conditional branches} triggers the generation of new block
versions. This is particularly useful since all type tests are conditional
branches. Versions are generated according to the types that actually occur
at run time. This {\em lazy code generation} approach has four key benefits:

\begin{enumerate}
\item The order in which versions for different type combinations are
generated tends to approximate the frequency of occurrence of the said types.
This is particularly helpful in the presence of a block version limit.

\item It tends to produce efficient, cache-friendly linear orderings of
compiled blocks in memory, as versions are generated in the order they are
first executed.

\item Neither memory nor time are wasted compiling block versions for type
combinations that never occur at run time. Type combinations that do not occur
are never accounted for.

\item Unexecuted blocks are never compiled. Exception handling code is not
generated for programs which do not throw exceptions. Floating point code is
not generated for programs which do not make use of floating point values.
\end{enumerate}

The Higgs backend lazily compiles versions of individual SSA basic blocks into x86-64
machine code as they are first executed. Higgs does not compile whole functions
at once. Instead, the JIT compilation model employed by
Higgs interleaves execution and compilation of basic blocks. The last
instruction of a block, which must be a branch instruction, determines which
block will be compiled next. If the branch is unconditional, or if its
direction can be determined at compilation time, no branch instruction is
generated, and the successor version the branch leads to is immediately
compiled (unless already compiled, in which case a direct jump is
written instead).

When a conditional branch whose direction cannot be determined at
compilation time is encountered, a pair of out-of-line stubs are generated
for the two possible outcomes of the branch, and execution resumes.
Stubs, when executed, call back the compiler requesting compilation of
the corresponding destination basic block with the typing context at
the branch.  The branch is then overwritten to fall through or jump to
the generated basic block version. This way, the compilation of a particular
basic block version is delayed until it is required for execution.

\subsection{Type Tags and Runtime Primitives\label{sec:type_tags}}

Higgs segregates values into a few categories based on
type tags~\cite{type_tags}. These categories are: 32-bit integers ({\tt int32}),
64-bit floating point values ({\tt float64}), miscellaneous JavaScript
constants ({\tt const}), and four kinds of garbage-collected pointers inside
the heap ({\tt string}, {\tt object}, {\tt array}, {\tt closure}). These
type tags form a simple, first-degree notion of types that is used to drive
the basic block versioning approach.

We chose this coarse-grained type classification to investigate
the effectiveness and potential of basic block versioning.
Higgs implements JavaScript operators as runtime library functions written in
an extended dialect of JavaScript, and most of these functions use type tags
to do type dispatching. As such, eliminating this first level of type tests
as well as boxing and unboxing overhead, is crucial to improving the
performance of the system as a whole.

\begin{figure}
\begin{lstlisting}[language=javascript]
function add(x, y) {
    if (is_i32(x)) { // If x is integer
        if (is_i32(y)) {
            if (var r = add_i32_ovf(x, y))
                return r;
            else // Handle the overflow case
                return add_f64(i32_to_f64(x),
                               i32_to_f64(y));
        } else if (is_f64(y))
            return add_f64(i32_to_f64(x), y);
    } else if (is_f64(x)) { // If x is fp
        if (is_i32(y))
            return add_f64(x, i32_to_f64(y));
        else if (is_f64(y))
            return add_f64(x, y);
    }

    // Eval args as strings and concat them
    return strcat(toString(x), toString(y));
}
\end{lstlisting}
\caption{Implementation of the {\tt +} operator\label{fig:rt_add}}
\end{figure}

Figure \ref{fig:rt_add} illustrates the implementation of the {\tt +} operator
as an example. As can be seen, this function makes extensive use of
low-level type test primitives such as {\tt is\_i32} and
{\tt is\_f64} to implement dynamic dispatch based on the type tags
of the input arguments. Most other arithmetic, comparison and property access
primitives implement a similar dispatch mechanism.

Note that while according to the ES5 specification all JavaScript numbers are
IEEE double-precision floating point values, high-performance JavaScript
VMs typically attempt to represent small integer values
using machine integers so as to improve performance by using
lower latency integer arithmetic instructions. We have made the same design
choice for Higgs. Consequently, JavaScript numbers are represented using tagged
{\tt int32} or {\tt float64} values.  Arithmetic operations on {\tt int32}
values may yield an {\tt int32} or {\tt float64} result, but arithmetic
operations on {\tt float64} values always yield an {\tt float64} result.

\subsection{Flow-based Representation Analysis}

To provide a point of comparison and contrast the capabilities of basic block
versioning with that of more traditional type analysis approaches, we have
implemented a forward flow-based representation analysis that computes a
fixed point on the types of SSA values. The analysis is an adaptation of
Wegbreit's algorithm as described in~\cite{sccp}. It is an intraprocedural
constant propagation analysis that propagates the types of SSA values in a
flow-sensitive manner.

The representation analysis uses sets of possible type tags as a type
representation. It is able to gain information from typed primitives
(e.g. {\tt add\_f64} produces {\tt float64} values) as well as type tests and
forward this information along branches. The analysis is also able to deduce,
in some cases, that specific branches will not be executed and ignore the
effects of code that was determined dead. The type tags are the same as those
used by basic block versioning, with the difference that basic block versioning
only propagates unique known types and not type sets (e.g. {\tt int32} $\cup$
{\tt float64}). This means that basic block versioning can only propagate
positive information gained from type tests whereas the analysis can propagate
both positive and negative information (e.g. {\tt a} is not {\tt int32}).

We have chosen to give the type analysis a richer type representation than
that of basic block versioning because several common arithmetic primitives
can produce overflows that cannot be statically predicted. This means that
most arithmetic operations can produce either {\tt int32} or {\tt float64}
types. If the type analysis could not represent this type set, it would be
forced to infer that the output type of most arithmetic operations is of
{\em unknown} type. This would immediately put the type analysis at an enormous
disadvantage when compared to basic block versioning because basic block
versioning is not affected by overflows that do not occur at run time.

\subsection{Concrete Example}

\begin{figure*}[!htb]
\begin{lstlisting}[language=javascript]
function sum(n) {
    for (var i=0, s=0; i<n; i++)
        s += i;
    return s;
}
\end{lstlisting}
\caption{The {\tt sum} function.\label{fig:sum}}
\end{figure*}

\begin{figure*}[!htb]
\begin{centeredbox}
\includegraphics[scale=0.47]{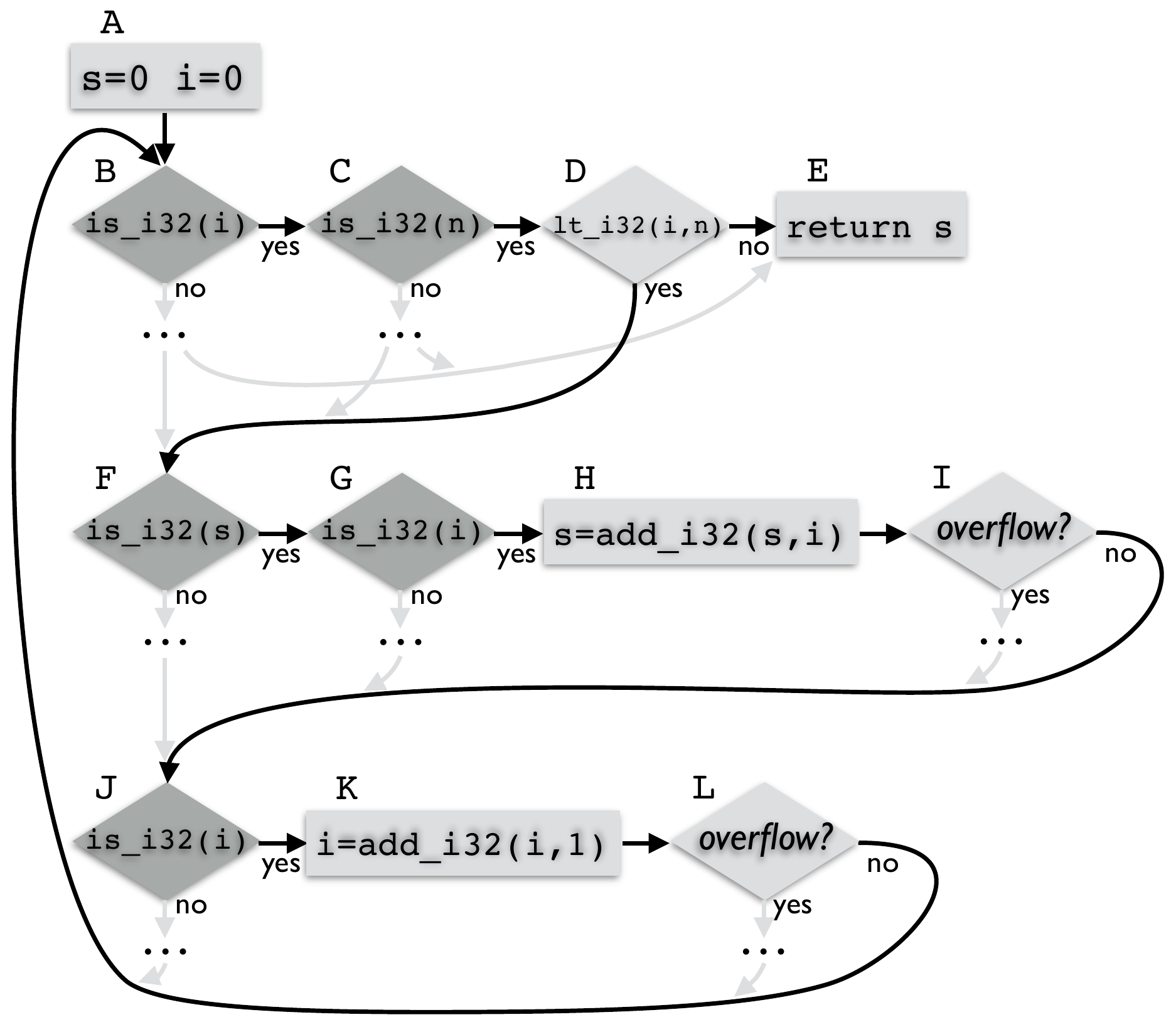}
\end{centeredbox}
\caption{Control flow graph of {\tt sum} function (unexecuted parts omitted).\label{fig:cfg-before-bbv}}
\end{figure*}

\begin{figure*}[!htb]
\begin{centeredbox}
\includegraphics[scale=0.47]{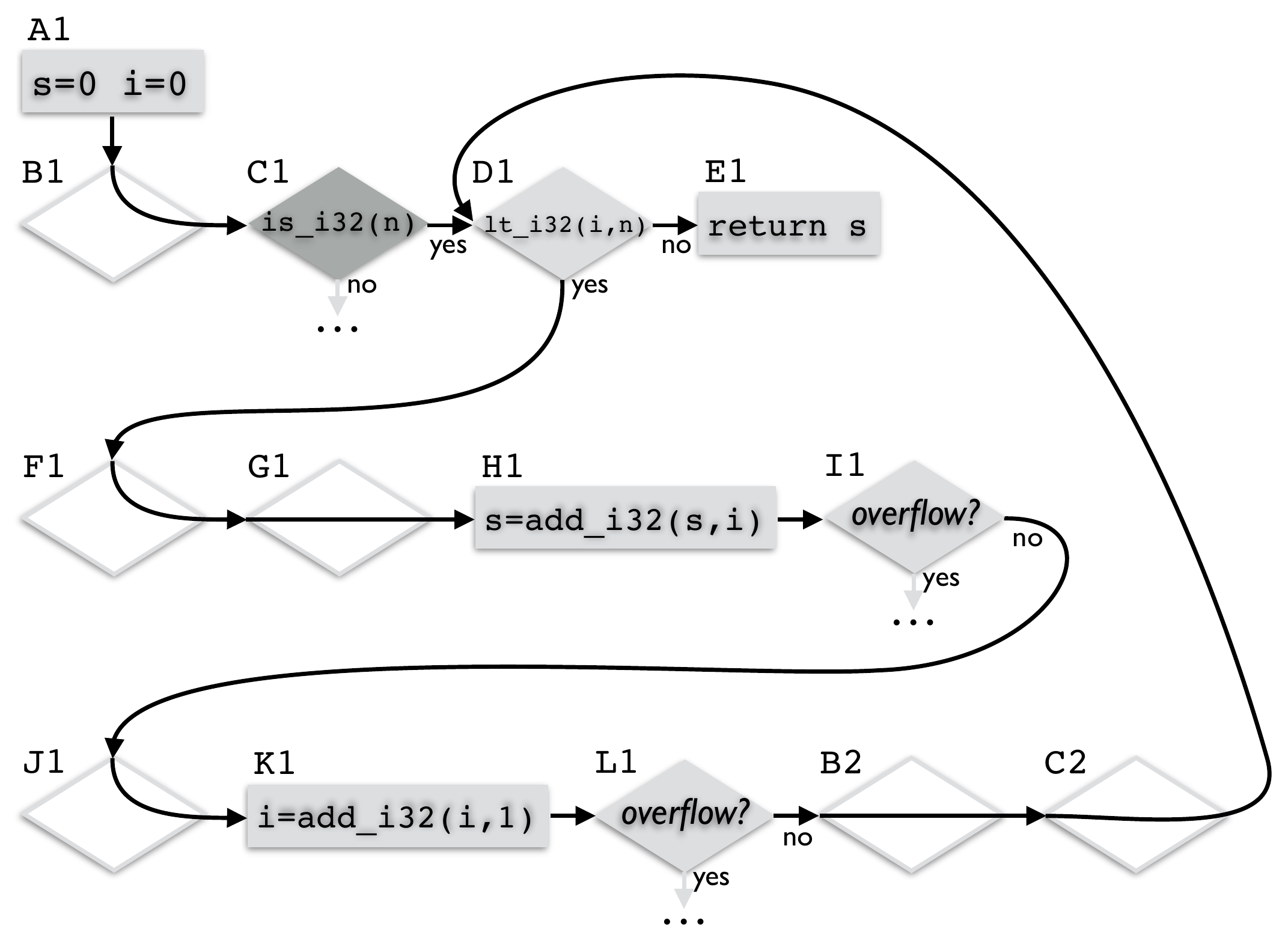}
\end{centeredbox}
\caption{Control flow graph of {\tt sum} function transformed by basic block versioning.\label{fig:cfg-after-bbv}}
\end{figure*}

\begin{figure*}[!htb]
\begin{centeredbox}
\includegraphics[scale=0.62]{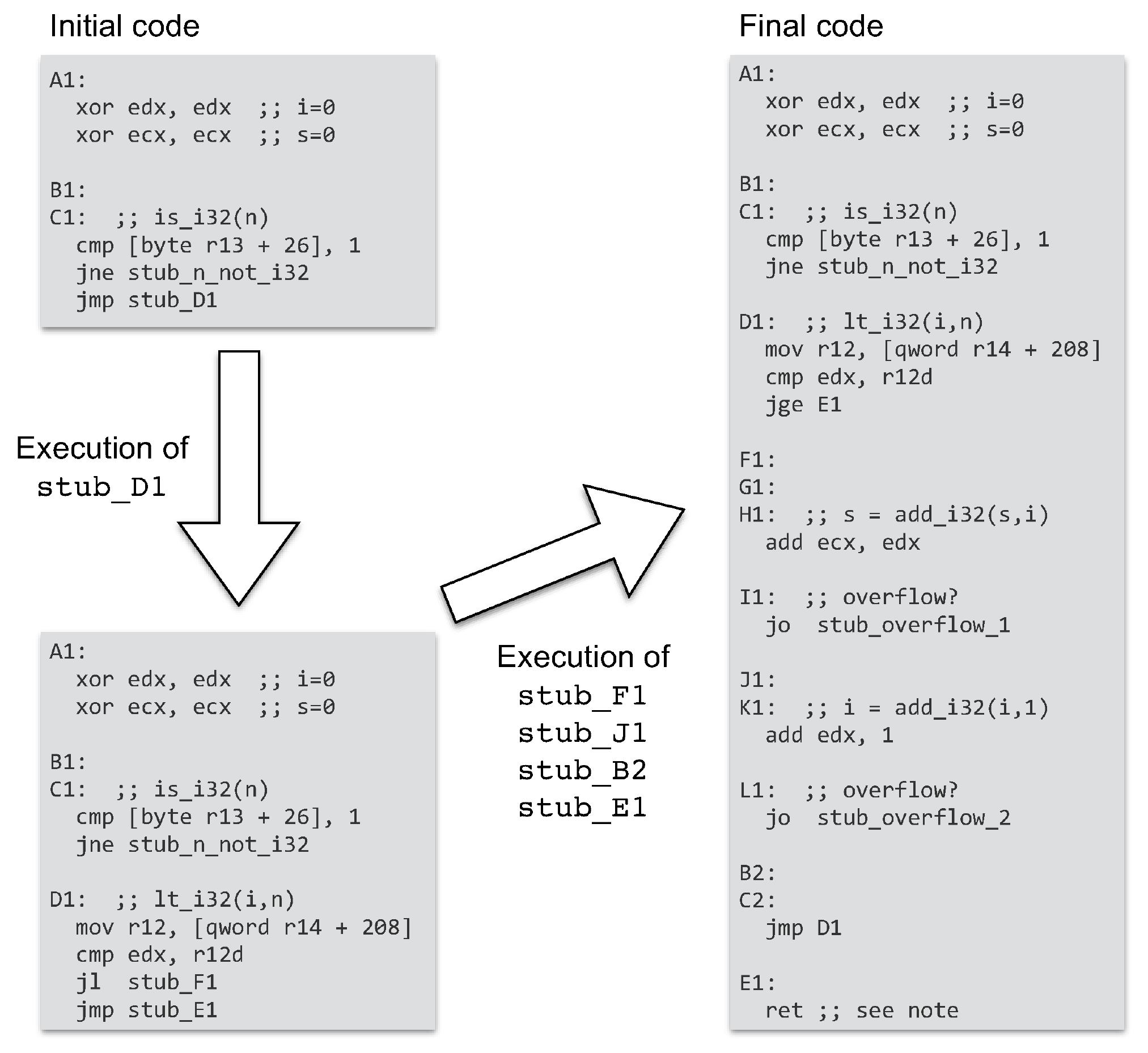}
\end{centeredbox}
\caption{Machine code at different steps of code compilation\label{fig:machine-code}}
\end{figure*}

To illustrate the lazy basic block versioning approach, we will
explain the compilation of the {\tt sum} function
given in Figure~\ref{fig:sum}.  Specifically, we
trace the execution of the function call {\tt sum(500)}. This only
requires 32-bit integer computations because no overflows occur for this
value of {\tt n}. Figure~\ref{fig:cfg-before-bbv} shows the parts of the
control flow graph of the function executed during this
call. The complete graph is larger and includes
code to handle floating point values and other types. Unexecuted
parts of the control flow graph are shown as ellipses (\ldots).

Before versioning, there are 5 type tests on {\tt i} and {\tt n} executed as
part of the loop. Higgs compiles the code for the {\tt sum} function incrementally,
type-specializing and eliminating type tests as compilation proceeds. The
compiled and specialized code is equivalent to the control flow graph shown
in Figure~\ref{fig:cfg-after-bbv}. Multiple blocks have been specialized based on the
knowledge that {\tt i}, {\tt s} and {\tt n} are of the {\tt int32} type.
Only one type test is left, in block {\tt C1}, and this type test has been
hoisted out of the loop. It is executed only once per call to {\tt sum}.

The incremental compilation process occurs in six steps and is illustrated in
Figure~\ref{fig:machine-code}. When first entering the {\tt sum} function, a version of
the entry block {\tt A} is compiled, generating {\tt A1}.
Variables {\tt s} and {\tt i} are initialized to {\tt int32} and this
is noted in the current typing context.  Then, block {\tt B} is compiled down to
nothing because {\tt i} is known to be {\tt int32} in the current context.
In {\tt C1}, generated from block {\tt C},
the type test on {\tt n} needs to emit machine code because the type of {\tt n}
is {\em unknown} in the current context and so must be tested. Therefore, stubs
{\tt stub\_n\_not\_i32} and {\tt stub\_D1} are generated and execution resumes at {\tt A1}.

Because {\tt n} contains an {\tt int32}, execution flows to {\tt stub\_D1},
which calls back into the JIT compiler. The branch instructions at the end of block {\tt C1} is
rewritten so that a jump to a stub is executed only if {\tt n} is not {\tt int32}.
In future calls of {\tt sum} where {\tt n} is {\tt int32}, the
branch will fall through to
block {\tt D1}. The generation of block {\tt D1} from {\tt D} is handled similarly.
Two stubs ({\tt stub\_F1} and {\tt stub\_E1}) are used to determine the direction of the less-than comparison
branch, which is unknown at compilation time.  Execution then resumes at {\tt D1} and flows to {\tt stub\_F1}.  This time, the JIT compiler inverts the direction of the branches at the end of block {\tt D1} so that the fall through will be block {\tt F1}.  Then blocks {\tt F1}, {\tt G1}, {\tt H1}, and {\tt I1} are generated and execution resumes at {\tt F1}.

The code is incrementally generated in this fashion by successively executing
{\tt stub\_J1}, {\tt stub\_B2}, and {\tt stub\_E1}. After the execution of
{\tt stub\_B2}, the
emitted code executes until the end of the loop. In the last loop iteration, the
less-than comparison in {\tt D1} fails. This triggers compilation of the loop
exit block {\tt E1}, which is conveniently placed outside of the loop body.
We note that the detailed sequence of instructions needed to return from
{\tt sum} is more complex than what is shown (to support JavaScript's variable
arity function calls).

The right part of Figure~\ref{fig:machine-code}
shows the generated code after the execution of {\tt sum(500)}
has completed. Type tests in blocks {\tt F1}, {\tt G1} and {\tt J1} were
eliminated because {\tt i}, {\tt s} and {\tt n} are known to be
{\tt int32} at those points. The jump back to the loop header in
{\tt L1} generated new versions of blocks {\tt B} and {\tt C} where
{\tt i}, {\tt s} and {\tt n} are known to be {\tt int32}. Hence, only the
first loop iteration performs a type test.

%% file: evaluation.tex
\subsection{Experimental Setup}

To assess the effectiveness of basic block versioning, we have tested it on a
total of 26 classic benchmarks from the SunSpider and Google V8 suites. One
benchmark from the SunSpider suite and one from the V8 suite were not included
in our tests because Higgs does not yet implement
the required features. Benchmarks making use of regular expressions were
discarded because unlike V8 and TraceMonkey, Higgs does not implement JIT
compilation of regular expressions, and neither does Truffle
JS~\cite{trufflejs, truffle}.

Since Higgs interleaves compilation and execution and which parts of a
program are eventually compiled is entirely dependent on run time behavior,
we have measured approximate compilation times using a microsecond counter
which is started and stopped when compilation begins and ends. The total
times accumulated are averaged across 10 runs to give a final compilation
time figure.

To measure execution time separately from compilation time in a manner
compatible with V8, TraceMonkey, Truffle JS and Higgs, we have modified
benchmarks so that they could be run in a loop. A number of warmup iterations
are first performed so as to trigger JIT compilation and optimization of code
before timing runs take place.

The number of warmup and timing iterations were scaled so that short-running
benchmarks would execute for at least 1000ms in total during both warmup and
timing. Unless otherwise specified, all benchmarks were run for at least 10
warmup iterations and 10 timing iterations.

V8 version 3.29.66, TraceMonkey version 1.8.5+ and Truffle JS v0.5 were used
for performance comparisons. Tests were executed on a system equipped with an
Intel Core i7-4771 quad-core CPU with 8MB L3 cache and 16GB of RAM running
Ubuntu Linux 12.04. Dynamic CPU frequency scaling was disabled for our
experiments.

\subsection{Dynamic Type Tests}

\begin{figure*}[!htb]
\begin{centeredbox}
\includegraphics[scale=1.00]{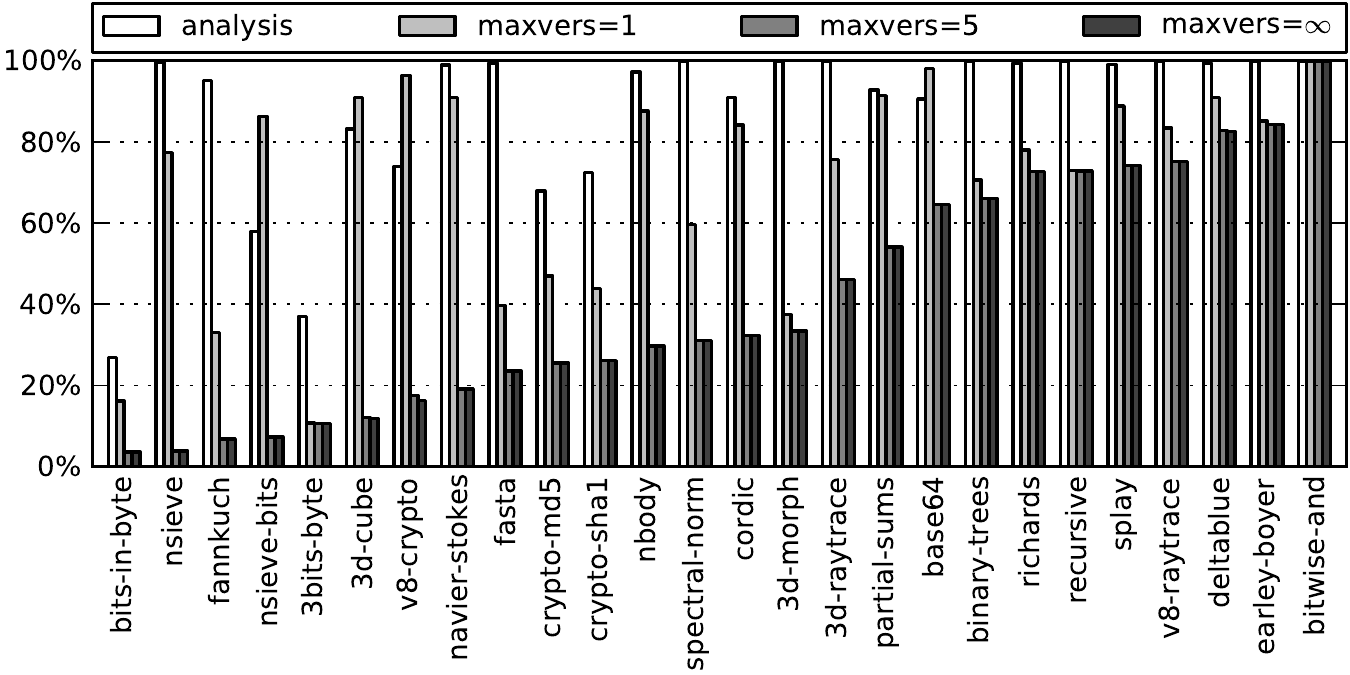}
\end{centeredbox}
\caption{Dynamic counts of type tests executed using the representation analysis and lazy basic block versioning with various version limits (relative to baseline)\label{fig:test_counts}}
\end{figure*}

Figure \ref{fig:test_counts} shows the dynamic counts of type tests
for the representation analysis and for lazy basic block
versioning with various block version limits. These counts are relative to
a baseline which has the version limit set to 0, and thus only generates a
generic version of each basic block.
As can be seen from the counts, the
analysis produces a reduction in the number of dynamically executed type
tests over the unoptimized baseline on most benchmarks. The basic block
versioning approach does at least as well as the analysis, and almost always
significantly better. Surprisingly, even with a version cap as low as 1
version per basic block, the versioning approach is often competitive with the
representation analysis. This is likely because most value types are
monomorphic.

\begin{figure}[!htb]
\begin{lstlisting}[language=javascript]
function bitsinbyte(b) {
    var m = 1, c = 0;
    while(m < 0x100) {
        if(b & m) c++;
        m <<= 1;
    }
    return c;
}

function TimeFunc(func) {
    var x, y, t;
    for(var x=0; x<350; x++)
        for(var y=0; y<256; y++) func(y);
}

TimeFunc(bitsinbyte);

\end{lstlisting}
\caption{SunSpider bits-in-byte benchmark\label{fig:bits-byte}}
\end{figure}

Raising the version cap reduces the number of type tests performed with the
versioning approach in an asymptotic manner as we get closer to the
limit of what is achievable with our implementation. The versioning
approach does quite well on the {\tt bits-in-byte} benchmark. This benchmark
(see Figure \ref{fig:bits-byte}) is an ideal use case for our versioning
approach. It is a tight loop performing bitwise and arithmetic
operations on integers which are all stored in local variables. The versioning
approach performs noticeably better than the analysis on this test because it
is able to test the type of the function parameter {\tt b}, which is initially
unknown when entering {\tt bitsinbyte} only once per function call and
propagate this type thereafter. The analysis on its own cannot achieve this,
and so must repeat the test for each operation on {\tt b}. In contrast,
the {\tt bitwise-and} benchmark operates exclusively on global variables, for
which our system cannot extract types, and so neither the type analysis nor
the versioning approach show any improvement over baseline for this benchmark.

\begin{figure}[!htb]
\begin{centeredbox}
\includegraphics[scale=1.00]{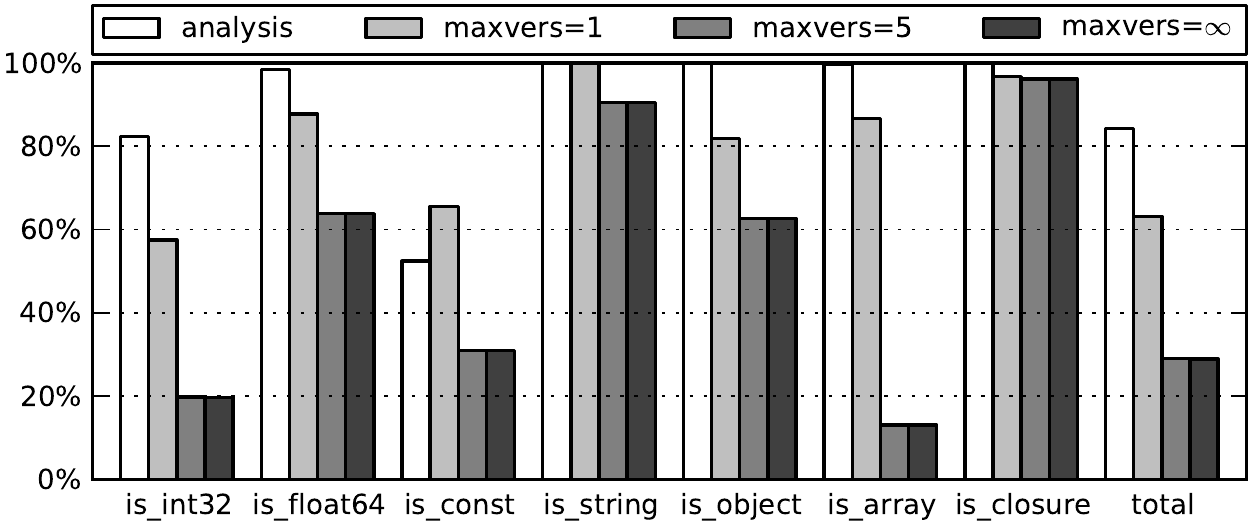}
\end{centeredbox}
\caption{Type test counts by kind of type test (relative to baseline)\label{fig:test_kinds}}
\end{figure}

A breakdown of relative type test counts by kind, averaged across all
benchmarks (using the geometric mean) is shown in Figure \ref{fig:test_kinds}.
We see that the versioning approach is able to perform as well or better than
the representation analysis across each kind of type test. The {\tt is\_closure}
category shows the least improvement. This is because functions are
typically globals or methods, which basic block versioning cannot yet
get type information about. We note that versioning is much more effective
than the analysis when it comes to eliminating {\tt is\_i32} type tests.
This is because integer and floating point types often get intermixed,
leading to cases where the analysis cannot eliminate such tests. The
versioning approach has the advantage that it can replicate and detangle
integer and floating point code paths. A limit of 5 versions per block
eliminates \input{testredumv5}\unskip\% of total type tests,
compared to \input{testredutpa}\unskip\% for the
analysis.

\subsection{Code Size Growth}

\begin{figure}[!htb]
\begin{centeredbox}
\includegraphics[scale=1.00]{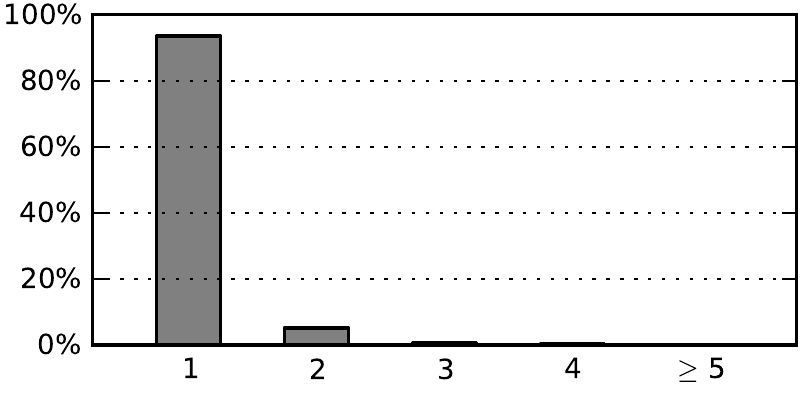}
\end{centeredbox}
\caption{Relative occurrence of block version counts\label{fig:ver_counts}}
\end{figure}

Figure \ref{fig:ver_counts} shows the relative proportion of blocks for which
different counts of versions were generated across all benchmarks. As one
might expect, the relative proportion of blocks tends
to steadily decrease as the number of versions is increased. Most basic
blocks have only one version, \input{perc2vers}\unskip\% have two, and just
\input{perc5vers}\unskip\% of blocks have 5 versions or more.
Hence, blocks with a large number of versions are a rare occurrence.

The maximum number of versions ever produced for a given block across our
benchmarks is \input{maxverscnt}\unskip. This occurs in the
{\tt \input{maxversbench}\unskip} benchmark. The function generating the most
block versions in this benchmark is {\tt rayTrace}.
This function is at the core of the ray tracing algorithm. 
It contains a loop with several live variables used during
iteration. Some of these variables can be either {\tt null} or an object
reference. There are also versions generated where basic block versioning
cannot determine a type for some variables.

\begin{figure*}[!htb]
\begin{centeredbox}
\includegraphics[scale=1.00]{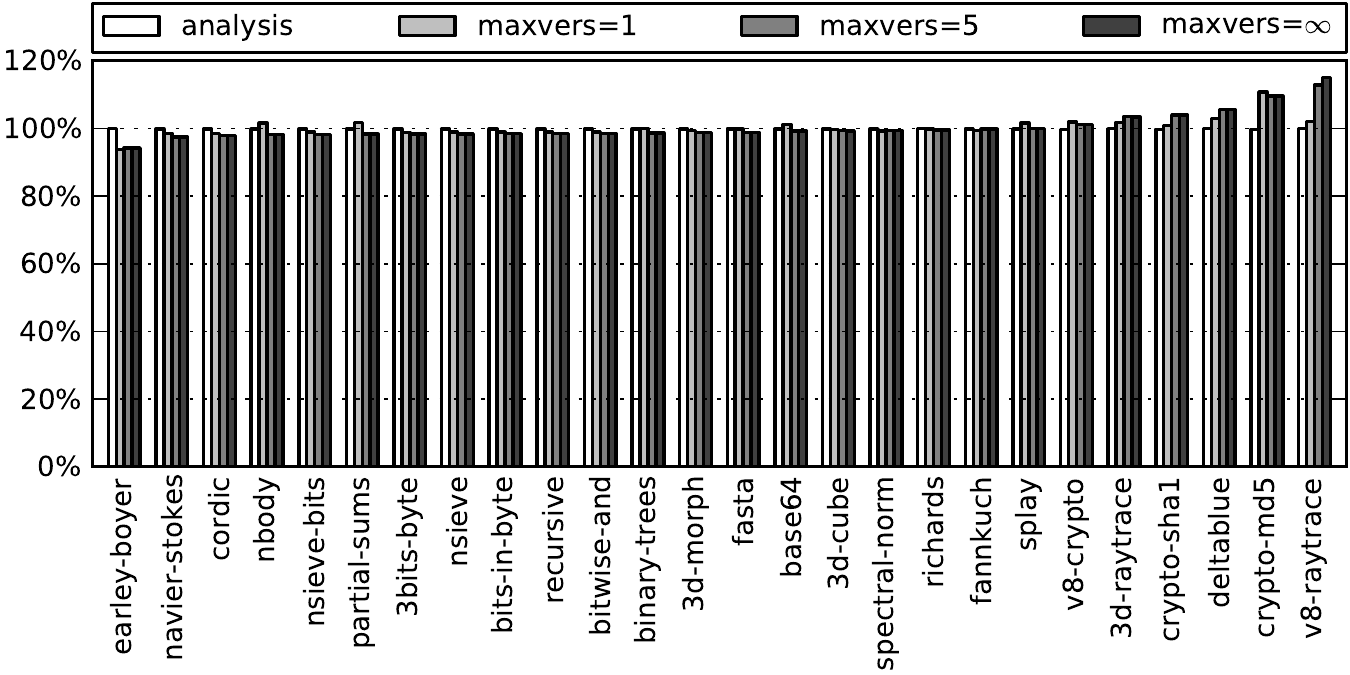}
\end{centeredbox}
\caption{Code size for various block version limits (relative to baseline)\label{fig:code_size}}
\end{figure*}

The effects of basic block versioning on the total generated code size are
shown in Figure \ref{fig:code_size}. It is interesting to note that the
representation analysis almost always results in a slight reduction in code size.
This is because the analysis allows the elimination of type tests and the
generation of more optimized code, which is usually smaller. On the other hand,
basic block versioning can generate multiple versions of basic blocks,
which often (but not always) results in more generated code. The volume of
generated code does not increase linearly with the block version limit. Rather,
it tapers off as a limited number of versions tends to be generated for each
block. A limit of 5 versions per block results in a mean code size increase of
\input{codeincrmv5}\unskip\%. With no limit at all on the number of versions,
the code size increase does not change much, with
a mean of \input{codeincrmvi}\unskip\% and a maximum increase of
\input{maxcodeincrmvi}\unskip\% across all benchmarks. On the
benchmarks we have tested, there is no pathological code size explosion, and
the block version limit is not strictly necessary.

\subsection{Execution Time}

\begin{figure*}[!htb]
\begin{centeredbox}
\includegraphics[scale=1.00]{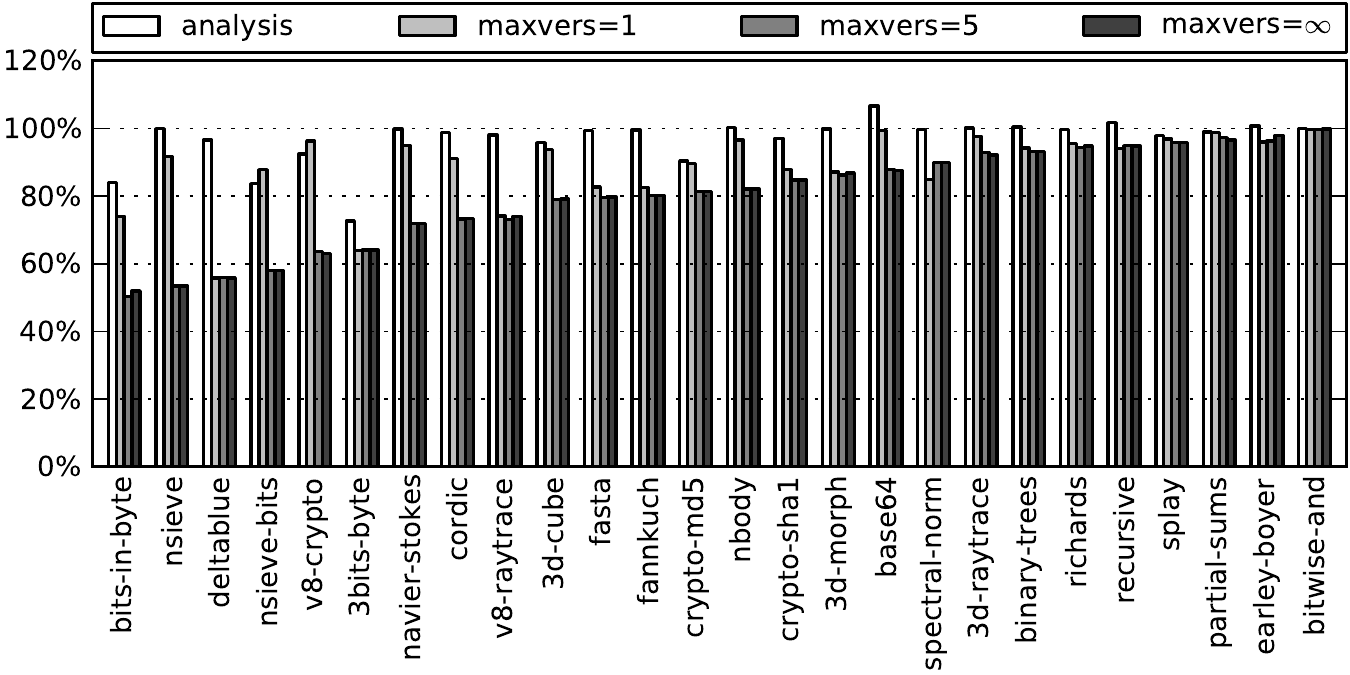}
\end{centeredbox}
\caption{Execution time for various block version limits (relative to baseline)\label{fig:exec_time}}
\end{figure*}

Figure \ref{fig:exec_time} shows the execution times relative to
baseline. Because our type analysis is not optimized for speed and incurs a
significant compilation time penalty, we have excluded compilation time and
measured only time spent executing compiled machine code. A limit of 5 versions
per block produces on average a \input{meanspeedupmv5}\unskip\%
reduction in execution time, and speedups of up to
\input{maxspeedupmv5}\unskip\%, while the type analysis yields a
\input{meanspeeduptpa}\unskip\% average speedup.

In most cases, basic block versioning produces a notable reduction in relative
execution time that compares favorably with the static analysis. The
intraprocedural type analysis does not eliminate enough type tests to be
effective in improving execution times. We believe that it should be possible
to significantly improve upon the basic block versioning results with method
inlining and better optimized property accesses, which would expose more type
tests and more precise type information.

\subsection{Eager Versioning}

\begin{figure*}[!htb]
\begin{centeredbox}
\includegraphics[scale=1.00]{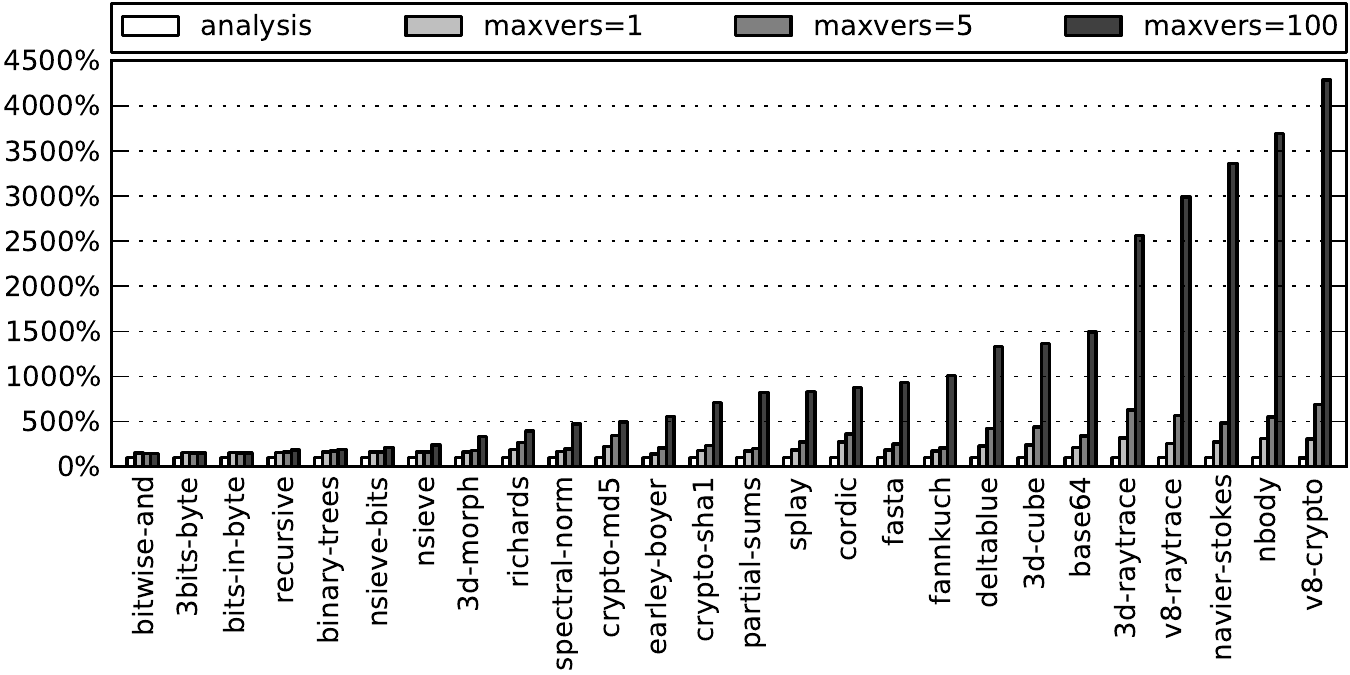}
\end{centeredbox}
\caption{Code size with eager basic block versioning (relative to baseline)\label{fig:eg_code_size}}
\end{figure*}

In order to evaluate the importance of lazyness in our basic block versioning
approach, we have tested an older version of Higgs which generates block
versions eagerly. In this configuration, whole methods are compiled at once,
never producing stubs, and specialized versions are generated for a given
block until the block version limit is hit. The versions are generated in no
particular order. The performance obtained with eager generation of block
versions was found to be inferior on all metrics. When the version limit is set to 5, on average,
the eager approach eliminates about half as many type tests as the lazy
approach, the code size is 223\% of baseline on average (see
Figure \ref{fig:eg_code_size}), and the execution time is 5\% slower than
baseline.

There are multiple issues with the eager generation of block versions. The
most important one is that without some form of lazyness, without code stubs,
we must always produce code for both sides of a conditional branch. In the case
of eager basic block versioning, this means we generate code for both
branches of a type test, even though in most cases only one side of the branch
is ever taken. We end up generating versions for a large number of type
combinations which cannot occur at run time, but which we have no heuristic to
discard at method compilation time. The number of possible type combinations
increases exponentially with the number of live variables, and so the block
version limit is rapidly reached. Since versions are generated in no particular
order, the specialized versions eagerly generated before the block version
limit is hit are likely to be versions matching irrelevant type combinations.

\subsection{Compilation Time}

\begin{figure*}[!htb]
\begin{centeredbox}
\includegraphics[scale=1.00]{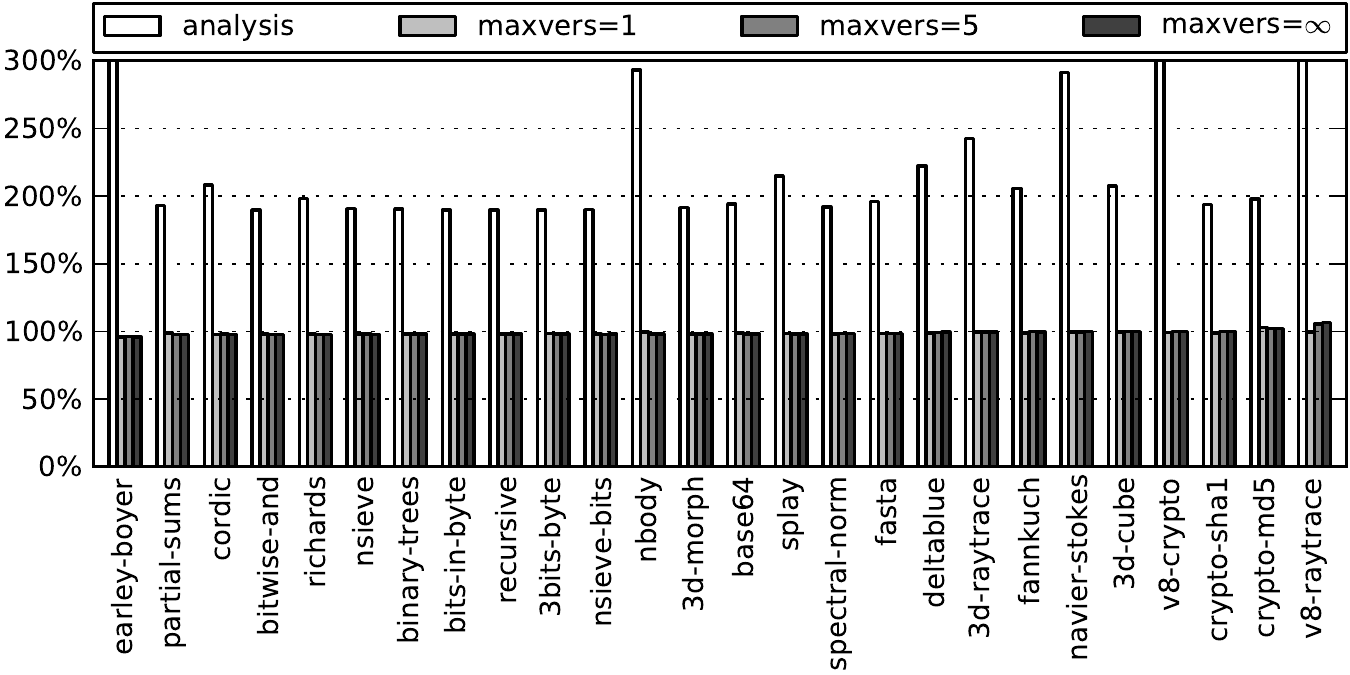}
\end{centeredbox}
\caption{Compilation time for various block version limits (relative to baseline)\label{fig:comp_time}}
\end{figure*}

The graph in Figure~\ref{fig:comp_time} shows a comparison of the total
compilation time with the type analysis and with different block version limits
relative to baseline. The type analysis, as implemented, is not particularly
efficient because it passes around maps of SSA values to type sets and iterates
until a fixed point is reached. This is expensive and scales poorly with
program size. The analysis increases compilation time by over 100\% in many
cases. In the worst case, on the {\tt earley-boyer} benchmark, the analysis
incurs a compilation time slowdown of more than 100 times.

Basic block versioning does not increase compilation times by much. A limit of
5 versions per block produces a compilation time decrease of
\input{meancompmv5}\unskip\% on average, and a
\input{maxcompmv5}\unskip\% increase in the worst case. It is
interesting to note that in many cases, enabling basic block versioning
reduces compilation time by a small amount. This is because specializing code
to eliminate type checks often makes it smaller, and for some basic blocks,
no machine code is generated at all.

\subsection{Comparison against the V8 Baseline Compiler}

\begin{figure*}[!htb]
\begin{centeredbox}
\includegraphics[scale=1.00]{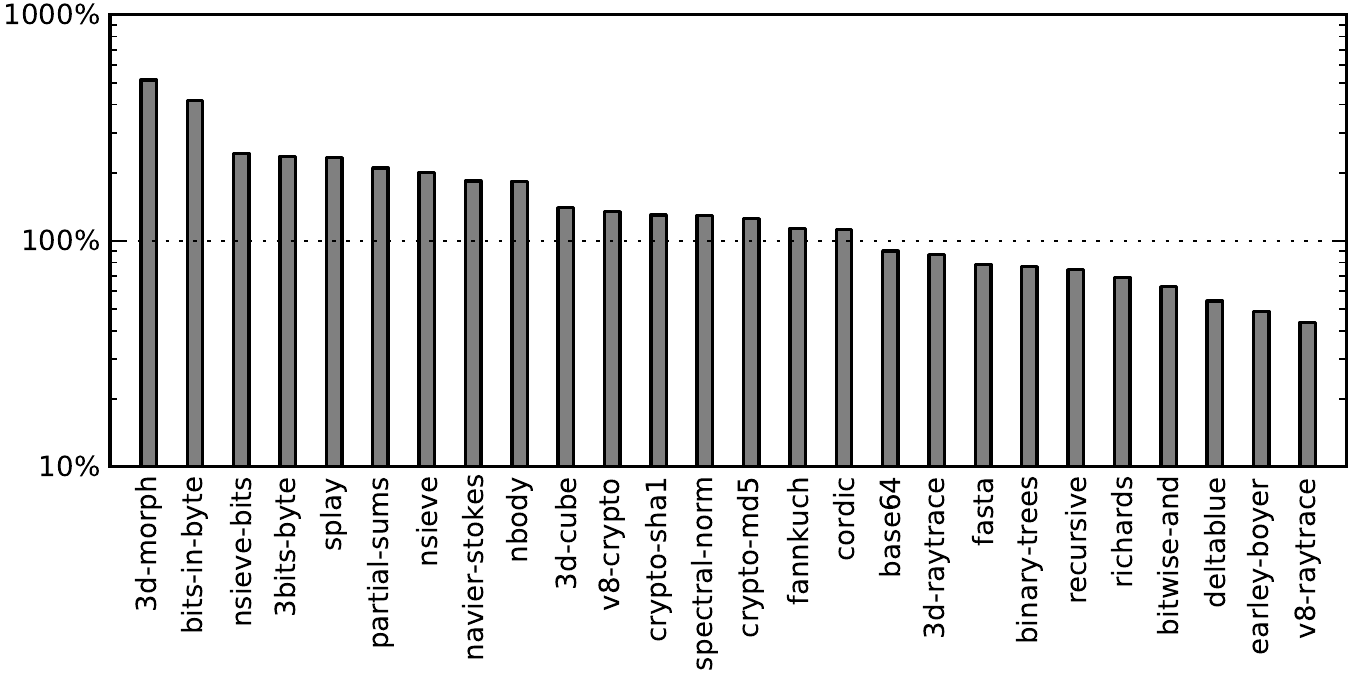}
\end{centeredbox}
\caption{Speedup relative to V8 baseline (log scale, higher is better)\label{fig:cmp_vs_v8}}
\end{figure*}

We have compared the execution time of the machine code generated by Higgs to
that of the V8 baseline compiler. The V8 baseline compiler is not to be
confused with Crankshaft. It is a low-overhead method-based JIT which, like
Higgs, does not perform method inlining and only performs basic optimizations
and fast on-the-fly register allocation. It is meant to compile code rapidly.

Figure~\ref{fig:cmp_vs_v8} shows speedups of Higgs over V8 baseline.
The scale is logarithmic, and higher bars indicate better performance on the part
of Higgs. As can be seen, Higgs delivers better performance on more than half of
the benchmarks. The three benchmarks on which V8 baseline does best are from the V8
suite, which the V8 baseline compiler was tailored to perform best on.
Higgs is able to deliver impressive speedups on a variety of benchmarks in
various areas of interest including floating point arithmetic, object-oriented
data structures and string manipulation.

\subsection{Comparison against TraceMonkey}

\begin{figure*}[!htb]
\begin{centeredbox}
\includegraphics[scale=1.00]{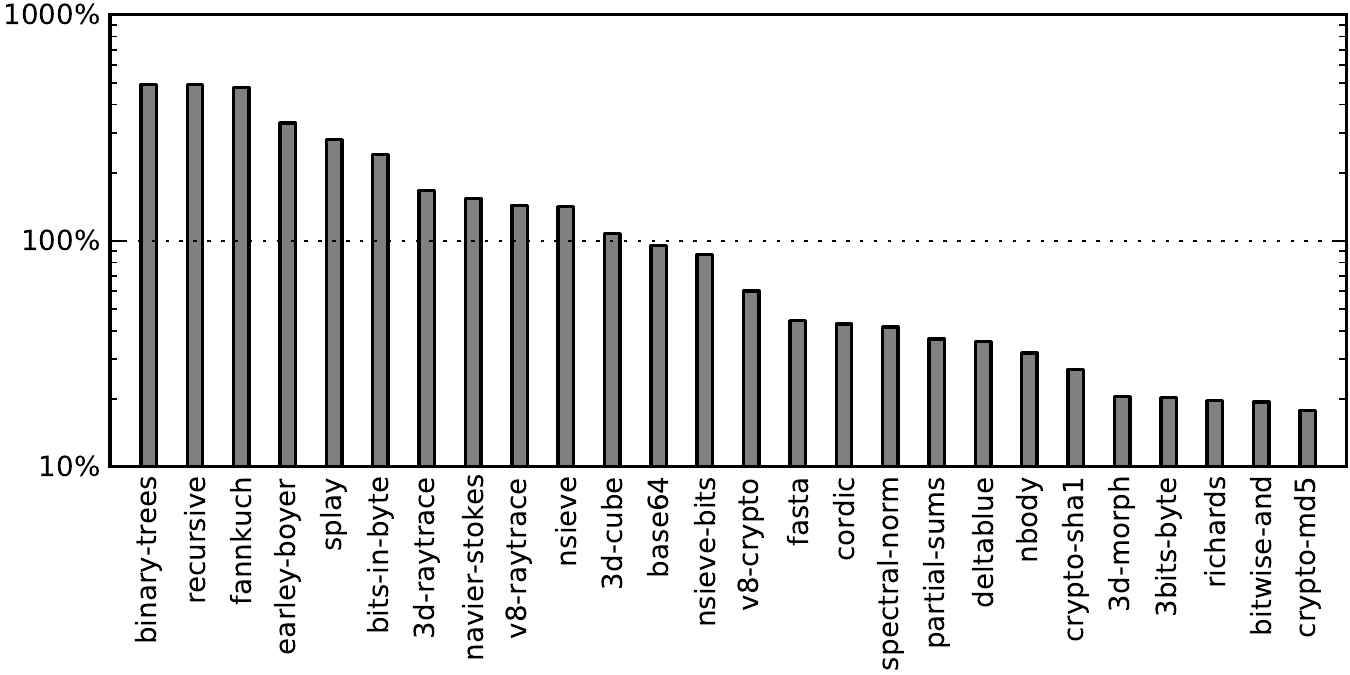}
\end{centeredbox}
\caption{Speedup relative to TraceMonkey (log scale, higher is better)\label{fig:cmp_vs_tm}}
\end{figure*}

The similarity of trace compilation and basic block versioning has prompted us
to compare Higgs to TraceMonkey, a tracing JIT compiler for
JavaScript that was part of Mozilla's SpiderMonkey until mid 2011.
It has the ability to eliminate type checks~\cite{trace_spec} based on analysis of traces.
Note that Higgs does not yet implement inlining of method calls whereas TraceMonkey can
inline them as part of tracing.

Figure~\ref{fig:cmp_vs_tm} shows speedups of Higgs over TraceMonkey. The scale
is again logarithmic, with higher bars indicating better performance on the part
of Higgs. TraceMonkey performs better on many benchmarks. Unsurprisingly, the
benchmarks TraceMonkey achieves the best performance on tend to be benchmarks
which include short and predictable loops. In these, TraceMonkey is
presumably able to inline all function calls which puts Higgs, without inlining,
at a significant performance disadvantage.

It is interesting that Higgs, even without inlining, does much
better on some of the largest benchmarks from our set. The two raytrace
benchmarks, for example, make significant use of object-oriented polymorphism
and feature highly unpredictable conditional branches. The {\tt earley-boyer}
benchmark is the largest of all and features complex control-flow. The
{\tt splay} and {\tt binary-trees} benchmarks apply recursive operations to
tree data structures. We note that Higgs performs much better than
TraceMonkey on the {\tt recursive} microbenchmark which suggests
TraceMonkey handles recursion poorly.

Higgs shines in benchmarks with complex, unpredictable control flow as well
as recursive computations. TraceMonkey is in no way the pinnacle of tracing
JIT technology, but there are clearly areas where basic block versioning
unambiguously wins over this implementation of trace compilation. Whereas
tracing, in its simplest forms, is ideal for predictable loops, basic block
versioning is not biased for any kind of control-flow structures. We believe
that implementing inlining in Higgs would likely even the performance gap on
the benchmarks where Higgs currently performs worse.

\subsection{Comparison against Truffle JS}

\begin{figure*}[!htb]
\begin{centeredbox}
\includegraphics[scale=1.00]{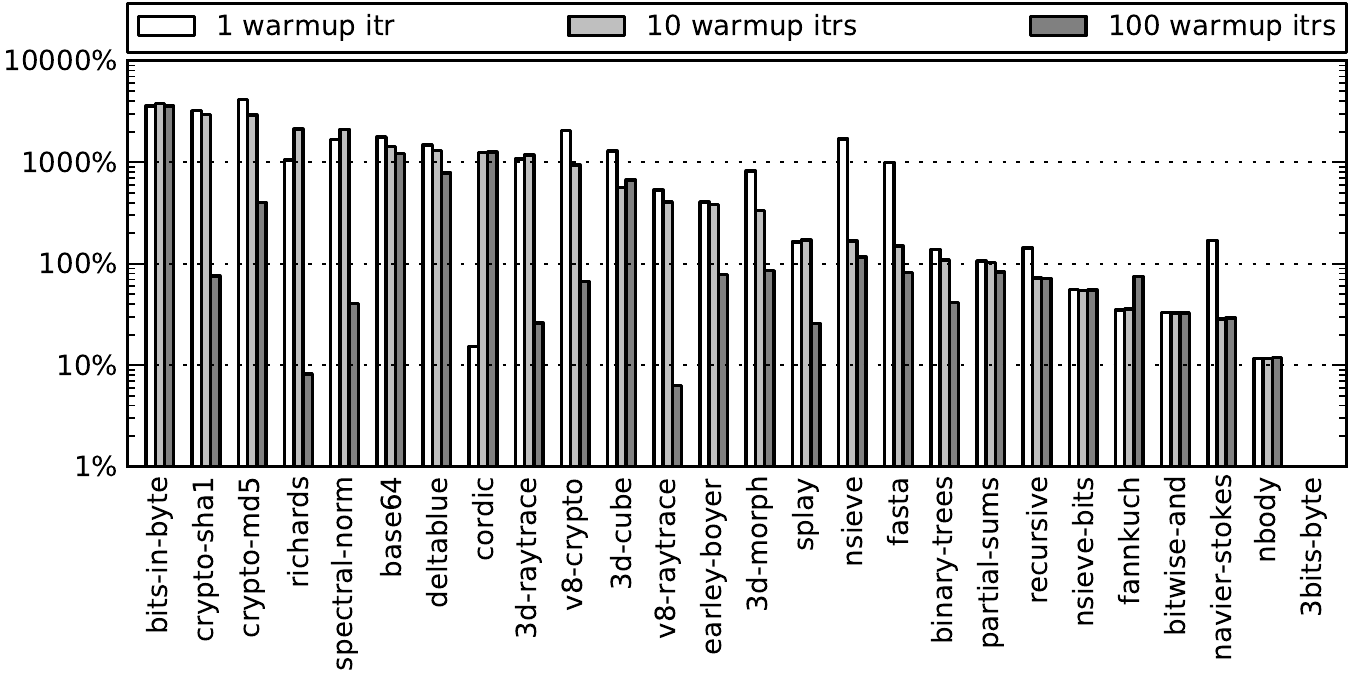}
\end{centeredbox}
\caption{Speedup relative to Truffle JS (log scale, higher is better)\label{fig:cmp_vs_tr}}
\end{figure*}

Figure~\ref{fig:cmp_vs_tr} shows the relative speed of Higgs over Truffle JS on a
logarithmic scale, with higher bars indicating better performance on the part
of Higgs. We have evaluated the performance with 1, 10 and 100 warmup iterations.
With 1 or 10 warmup iterations, Higgs outperforms Truffle on the majority of
benchmarks, with speedups of up to 30x in some cases.

With 100 warmup iterations, the picture changes, and Truffle outperforms Higgs
on most benchmarks. This seems to be because Truffle interprets code for a
long time before compiling and optimizing it. In contrast, Higgs only needs to
compile and execute a given code path once before it is optimized, with no
warmup executions required.

Truffle has two main performance advantages over Higgs. The first is
that after warmup, Truffle is able to perform deep inlining, as illustrated by
the {\tt v8-raytrace} benchmark. The second is that Truffle has sophisticated
analyses which Higgs does not have. For instance, the recorded time for the
{\tt 3bits-byte} microbenchmark is zero, suggesting that Truffle was able to
entirely eliminate the computation performed as its output is never used. Doing
this requires a side-effect analysis which can cope with the semantic
complexities of JavaScript.

We note that even with 100 warmup iterations, and despite Truffle's powerful
optimization capabilities, there remain several benchmarks where Higgs
performs best, with speedups over 10x in some cases.

%% file: testredutpa.tex
16

%% file: perc2vers.tex
5.2

%% file: perc5vers.tex
0.16

%% file: maxverscnt.tex
11

%% file: maxversbench.tex
v8-raytrace

%% file: codeincrmv5.tex
0.19

%% file: codeincrmvi.tex
0.25

%% file: maxcodeincrmvi.tex
15

%% file: meanspeedupmv5.tex
21

%% file: meanspeeduptpa.tex
4

%% file: meancompmv5.tex
1.2

%% file: maxcompmv5.tex
5.3

%% file: related.tex
The {\em tracelet-based} approach used by Facebook's HipHop VM for PHP
(HHVM)~\cite{hiphopvm} bears
much similarity to our own. It is based on the JIT compilation of small code
regions (tracelets) which are single-entry multiple-exit basic blocks. Each
tracelet is type-specialized based on variable types observed at JIT
compilation time. Guards are inserted at the entry of tracelets to verify at
run time that the types observed are still valid for all future executions.
High-level instructions in tracelets are specialized based on the guarded
types. If these guards fail, new versions of tracelets are compiled based
on different type assumptions and chained to the failing guards.

There are three important differences between the HHVM approach and basic block
versioning. The first is that BBV does not insert dynamic guards but instead
exposes and exploits the underlying type checks that are part of the definition
of runtime primitives. HHVM cannot do this as it uses monolithic high-level
instructions to represent PHP primitives, whereas the Higgs primitives are
self-hosted and defined in an extended JavaScript dialect.
The second difference is that BBV propagates known types to successors and 
doesn't usually need to re-check the types of local variables. A third
important difference is
that HHVM uses an interpreter as fallback when too many tracelet versions
are generated. Higgs falls back to generic basic block versions which do not
make type assumptions but are still always JIT compiled for better performance.

{\em Trace compilation}, originally introduced by the
Dynamo~\cite{dynamo} native code optimization system, and later applied to
JIT compilation in HotpathVM~\cite{hotpathvm} aims to record long sequences
of instructions executed inside hot loops. Such linear sequences of
instructions often make optimization simpler. Type information can be
accumulated along traces and used to specialize code and remove type
tests~\cite{trace_spec}, overflow checks~\cite{trace_ovf} or unnecessary
allocations~\cite{trace_alloc}. Basic block versioning resembles tracing in
that context updating works on essentially linear code fragments and code is
optimized similarly to what may be done in a tracing JIT. Code is also
compiled lazily, as needed, without compiling whole functions at once.

The simplicity of basic block versioning is one of its main advantages.
It does not require external infrastructure such as an
interpreter to execute code or record traces. Trace compiler
implementations must deal with corner cases that do not appear with
basic block versioning. With trace compilation, there is the potential for
trace explosion if there is a large number of control flow paths going through
a loop. It is also not obvious how many times a loop should be recorded or
unrolled to maximize the elimination of type checks. This problem is
solved with basic block versioning since versioning is driven
by type information. Trace compilers must implement parameterizable policies
and mechanisms to deal with recursion, nested loops and potentially very long
traces that do not fit in instruction caches.

{\em Run time type feedback} uses profiling to gather type information at execution
time. This information is then used to optimize dynamic
dispatch~\cite{type_feedback}. There are similarities with
basic block versioning, which generates optimized code paths lazily based on
types occurring at run time. The two techniques are complementary.
Basic block versioning could be made more efficient by using
type profiling to reorder sequences of type tests in a type dispatch. Type feedback
could be augmented by using basic block versioning to generate multiple
optimized code paths. The Truffle framework uses run time type feedback
combined with guards to type-specialize AST nodes at run
time~\cite{trufflejs, truffle}.

There have been multiple efforts to devise type analyses for dynamic languages.
The Rapid Atomic Type Analysis (RATA)~\cite{rata} is an intraprocedural
flow-sensitive analysis based on abstract interpretation that aims to assign
unique types to each variable inside of a function. Attempts have
also been made to define formal semantics for a subset of dynamic languages
such as JavaScript~\cite{ti_js}, Ruby~\cite{ti_ruby} and Python~\cite{rpython},
sidestepping some of the complexity of these languages and making them more
amenable to traditional type inference techniques. There are
also flow-based interprocedural type analyses for JavaScript based on
sophisticated type lattices~\cite{tajs}\cite{tajs_lazy}\cite{type_ref}. Such
analyses are usable in the context of static code analysis, but take too long
to execute to be usable in VMs and do not deal with the complexities of
dynamic code loading.

More recently, work done by Brian Hackett et al. at Mozilla resulted in an
interprocedural hybrid type analysis for JavaScript suitable for use in
production JIT compilers~\cite{mozti}. This analysis represents an important step
forward for dynamic languages, but as with other type analyses, must
conservatively assign one type to each value, making it vulnerable to
imprecise type information polluting analysis results. Basic block versioning
can help improve on the results of such an analysis by hoisting
tests out of loops and generating multiple optimized code paths where
appropriate.

Basic block versioning bears some similarities to classic compiler
optimizations such as {\em loop unrolling}~\cite{loop_unrolling}, {\em loop
peeling}~\cite{loop_peeling}, and {\em tail duplication}, considering it achieves
some of the same results. Another parallel can be drawn with
{\em Partial Redundancy Elimination (PRE)}~\cite{pre}; the versioning approach
seeks to eliminate and hoist out of loops a specific
kind of redundant computation: dynamic type tests. {\em Code
replication} has also been used to improve the effectiveness of
PRE~\cite{pre_repl}.

Basic block versioning is also similar to the idea of {\em node
splitting}~\cite{node_splitting}. This technique is an analysis device designed
to make control flow graphs reducible and more amenable to analysis.
The {\em path splitting} algorithm implemented in the SUIF compiler~\cite{path_splitting}
aims at improving reaching definition information by replicating control flow
nodes in loops to eliminate joins. Unlike basic block versioning, these
algorithms cannot gain information from type tests. The algorithms as presented
are also specifically targeted at loops, while basic block versioning makes no
special distinction. Mueller and Whalley have developed effective static
analyses that use {\em code replication} to eliminate both unconditional and
conditional branches~\cite{code_repl_uncond}\cite{code_replication}. However,
their approach is intended to optimize loops and operates on a low-level
intermediate representation that is not ideally suited to the
elimination of type tests in a high-level dynamic language.

{\em Customization} is a technique developed to optimize Self
programs~\cite{self_customization} that compiles multiple copies of methods
specialized on the receiver object type. Similarly, {\em type-directed
cloning}~\cite{type_cloning} clones methods based on argument types,
producing more specialized code using richer type information. The work of 
Chevalier-Boisvert et al. on {\em Just-in-time specialization} for
MATLAB~\cite{mcvm} and similar work done for the MaJIC MATLAB
compiler~\cite{majic_matlab} tries to capture argument types to dynamically
compile optimized versions of whole functions. All of these techniques are
forms of type-driven code duplication aimed at extracting type information.
Basic block versioning operates at a lower level of granularity, allowing
it to find optimization opportunities inside of method bodies by duplicating
code paths.

Basic block versioning also resembles the {\em iterative type analysis} and
{\em extended message splitting} techniques developed for Self by Craig
Chambers and David
Ungar~\cite{itr_analysis}. This is a combined static analysis and
transformation that compiles multiple versions of loops and duplicates 
control flow paths to eliminate type tests. The analysis works in an iterative
fashion, transforming the control flow graph of a function while performing
a type analysis. It integrates a mechanism to generate new versions of loops
when needed, and a message splitting algorithm to try and minimize type
information lost through control flow merges. One key disadvantage
is that statically cloning code requires being conservative,
generating potentially more code than necessary, as it is impossible to
statically determine exactly which control flow paths will be taken at run
time, and this must be overapproximated. Basic block versioning is simpler to
implement and generates code lazily, requiring less compilation time and memory
overhead, making it more suitable for integration into a baseline JIT compiler.

%% file: future.tex


Since Higgs is a standalone JavaScript VM that is not integrated in a web
browser, we have tested it on out-of-browser benchmarks that are most relevant
to using JavaScript in the server-side space (like node.js\footnote{http://nodejs.org}).
We do not anticipate any issues with using
basic block versioning in a JavaScript VM integrated into a web browser, but
we have not done the integration required for such an experiment.
Basic block
versioning is suitable for optimizing dynamic
languages in general, not just
JavaScript web applications in particular.

Several extensions to basic block versioning are possible. For instance, we
have successfully extended it to perform overflow check elimination on loop
increments, but have kept this feature disabled to simplify the presentation in
this paper. Another interesting extension of basic block versioning
would be to propagate information about global variable types, object identity
and object property types. It may also be desirable to know the exact value of
some variables and object fields, particularly for values likely to remain
constant.

The implementation of lazy basic block versioning evaluated in this paper only tracks type
information intraprocedurally. It would be beneficial to apply basic block
versioning to function calls so that type information can propagate from caller
to callee. This would entail having multiple specialized entry points for
parameter types encountered at the call sites of a function. Similarly, call
continuation blocks (return points) could be versioned to allow information
about return value types to flow back to the caller.

%% file: conclusion.tex
We have described a simple approach to JIT compilation called lazy
basic block versioning. This technique combines code generation with type
propagation and code duplication to produce more optimized code through the accumulation of type
information during code generation. The versioning approach is able to perform
optimizations such as automatic hoisting of type tests and efficiently
detangles code paths along which multiple numerical types can occur. Our
experiments show that in most cases, basic block versioning eliminates
significantly more dynamic type tests than is possible using a traditional
flow-based type analysis. It eliminates up to \input{testredumv5}\unskip\% of
type tests on average with a limit of 5 versions per block, compared to
\input{testredutpa}\unskip\% for the analysis we have tested, and
never performs worse than such an analysis.

We have empirically demonstrated that although our implementation of basic
block versioning does increase code size in some cases, the resulting increase
is quite small and pathological code size explosions are unlikely to occur.
In our experiments, a limit of 5 versions per block results
in a mean code size increase of just \input{codeincrmv5}\unskip\%.
Our experiments with Higgs also indicate that lazy basic block versioning
improves performance up to \input{maxspeedupmv5}\unskip\% with a
limit of 5 versions per block. Finally, we have shown that Higgs
performs better than the V8 baseline compiler on most of our benchmarks,
and better than TraceMonkey on several of the more complex benchmarks in our
set.

Basic block versioning is a simple and practical technique that requires
little implementation effort and offers important advantages in JIT-compiled
environments where type analysis is often difficult and costly. Dynamic
languages, which perform a large number of dynamic type tests, stand to
benefit the most.

Higgs is open source and the code used in preparing this publication is
available on GitHub\footnote{https://github.com/higgsjs/Higgs/tree/ecoop2015}.